\documentclass[pre,aps,twocolumn,superscriptaddress,amssymb,amsmath]{revtex4-2}
\usepackage{graphicx}
\usepackage{hyperref}
\usepackage{appendix}
\usepackage{csquotes}
\usepackage{color, soul}
\usepackage{float}
\usepackage{nicefrac}
\usepackage{tikz}
\usepackage{xcolor}

\newcommand{\etal}{\textit{et~al.}}

\DeclareRobustCommand{\tikzagreeone}{%
	\begin{tikzpicture}[baseline=-0.6ex,scale=0.9]
		\draw[blue, thick] (0,0) -- (0.5,0);
		\fill[blue] (0,0) circle (2.8pt);
		\fill[blue] (0.5,0) circle (2.8pt);
	\end{tikzpicture}%
}

\DeclareRobustCommand{\tikzagreetwo}{
	\begin{tikzpicture}[baseline=-0.6ex,scale=0.9]
		\draw[blue, thick] (0,0) -- (0.5,0);
		\fill[red] (0,0) circle (2.8pt);
		\fill[red] (0.5,0) circle (2.8pt);
	\end{tikzpicture}
}

\begin{document} 
    \title{Persistent Imbalance in Open Networks with Coevolutionary dynamics }
    
    \author{S. Arab Mohammadi}
    \affiliation{Department of Physics, Shahid Beheshti University, Evin, Tehran 1983969411, Iran}
    
    \author{H. Jafari}
    \affiliation{Department of Electrical Engineering, Sharif University of Technology, P.O. Box 11365-9161, Tehran, Iran}
    
    \author{A. Kargaran}
    \email{amir.kargaran@ipm.ir}
    \affiliation{School of Biological Sciences, Institute for Research in Fundamental Sciences (IPM), 19395-5746, Tehran, Iran}
    
    \author{A. Hosseiny}
    \affiliation{Department of Physics, Shahid Beheshti University, Evin, Tehran 1983969411, Iran}
    
    \author{G. Reza Jafari}
    \email{g\_jafari@sbu.ac.ir}
    \affiliation{Department of Physics, Shahid Beheshti University, Evin, Tehran 1983969411, Iran}
    
    \date{\today}
    
    \begin{abstract}
        Societies are quintessential open systems, shaped by internal dynamics as well as external influences. The question is how these external influences alter the collective behavior and network dynamics. To answer this, we investigate coevolutionary balance dynamics in a system of independent and open networks. Here, the system consists of two interacting networks with directed (asymmetric) coupling: an independent network evolving autonomously and an open (dependent) network whose dynamics are influenced by the former. Using a mean-field framework, we demonstrate a transition temperature: below the transition temperature, the independent network reaches a state of structural balance, while the open network is destabilized by persistent imbalance states and enters a sustained imbalance phase. This coupling also induces a measurable upward shift in the transition temperature. Direct numerical simulations robustly confirm these analytical predictions.
    \end{abstract}
        
    \maketitle

    \section{Introduction}
    Using networks, fundamental models have been developed to describe social phenomena, where links, nodes, or both can govern the system’s dynamics. Structural balance theory  \cite{heider1946attitudes, cartwright1956, heider1958} and numerous extensions and applications in signed networks \cite{hart1974, Szell2010, rabbani2019, kargaran2025, Noudehi2022, Oloomi2023, allahyari2022, Moradimanesh2021, saberi2024} have focused on the influence of local interactions over the global network configuration. Opinion formation theory  \cite{castellano2009, galam1996, holme2006nonequilibrium, Gross2006} examines how individual beliefs evolve through local interactions, leading to collective patterns. Additionally, homophily theory \cite{mcpherson2001birds} addresses the tendency of individuals to form connections with similar others. Building on these concepts, coevolutionary models have attracted attention, as they simultaneously consider the dynamics of nodes and links. Previous studies largely ignore interactions between communities, although such interactions can significantly influence internal dynamics. To study such effects, we consider a system of two interacting networks with directed (asymmetric) coupling: an independent network evolving autonomously and an open (dependent) network whose dynamics are influenced by the former. In this context, “open” refers to the fact that the dynamical variables of the dependent network explicitly depend on variables of another network. Within this framework, interactions are coevolutionary, being symmetric within each network and asymmetric across the two networks.
    
    In a coevolutionary model, nodes and links evolve reciprocally and interdependently: changes in node states affect link dynamics, while the concurrently evolving network structure simultaneously feeds back on node behavior. In social contexts, this interaction can be conceptualized through two primary mechanisms. First, the process of opinion formation and evolution among individuals, which has been studied in foundational models such as the voter model \cite{sood2005voter}. Second, the tendency to connect with individuals holding similar views, known as homophily, whose critical role in shaping network structure and facilitating the spread of behavior has been extensively documented \cite{lee2017homophily, Soderberg2002}. Early studies investigated simplified versions of the interaction between opinions and networks. For instance, Holme and Newman \cite{holme2006nonequilibrium} analyzed phase transitions in networks with binary links and a limited set of opinions. Similarly, Castellano analyzed opinion dynamics and phase transitions \cite{castellano2000prl, castellano2003epl}. Saeedian \etal\cite{Saeedian2020Absorbing, Saeedian2024Modelling} extended these models to show how the joint dynamics of node and link states with network rewiring drives critical structural changes. Singh \etal\cite{Singh2014, Singh2016} demonstrated that convergence times in coevolving networks are highly sensitive to Heb's rule and that the presence of committed individuals can significantly alter the speed of consensus formation. In recent years, more advanced models have emerged, including the integrated frameworks examining homophily and structural balance \cite{gorski2020coevolution, gorski2020coevolution2, phamPRL2023, phamPNAS2022,kan2021adaptive}. In the other work \cite{kargaran2021}, researchers analyze a coevolutionary model defined on \emph{node--link--node triplets}. In this setup, each triplet consists of two nodes connected by a link, and the triplet's state depends on the combination of the node states and the sign of the link. Overall, it demonstrates that the interplay between link and node behavior can lead to community fragmentation or consensus under varying conditions.
    
    Real-world systems often experience influence from other interacting systems, such that the dynamics of one system can be affected by external factors. Despite significant advances, the explicit role of such influences has rarely been incorporated into coevolutionary modeling. To address this limitation, we adopt the same two-network framework introduced above, consisting of an independent network and a dependent (open) network with directed coupling. Each network is modeled as a fully connected system of nodes and links with discrete states, where the dynamics are governed by triplet interactions. Within this framework, one network evolves through its internal dynamics, while the other, in addition to its intrinsic dynamics, is directly influenced by the state of the independent network. Within this framework, the primary research questions we investigate are: (i) How does asymmetric, one-way influence between the networks reshape the coevolutionary dynamics? (ii) What is the nature of possible phase transitions under asymmetric coupling? (iii) How do the initial conditions of each network affect the long-term behavior and stability of the coevolutionary model? 
    
    To address these questions, in Section~\ref{model} we first adopt a non-equilibrium framework, with differential equations for nodes and links \cite{Singh2012}. Subsequently, motivated by our interest in the long-time behavior of the system, we reformulate the equations within an equilibrium Hamiltonian framework. To solve the proposed Hamiltonian and calculate the energy of the dependent network, we employ the mean-field approximation. Section~\ref{analysis} presents the mean-field solution and simulation results. For the analytical treatment, the Exponential Random Graph (ERG) framework \cite{park2004statistical,park2005solution,snijders2006ergm,robins2007ergm, lusher2012ergm, Krause2020} is utilized. The main findings are summarized in Section~\ref{sec:conclusion}, and technical details are provided in the appendices.

    \begin{figure}
        \centering
        \includegraphics[width=0.9\linewidth]{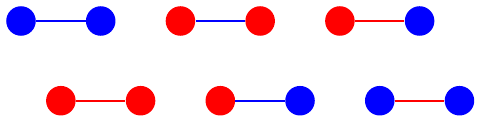}
        \caption{Balanced (top row) and imbalanced (bottom row) node-link–node triplets in the coevolutionary model. Nodes colored blue (red) represent positive (negative) opinions, while blue (red) links indicate agreement (disagreement).}
        \label{fig:main-fig1}
    \end{figure}
    
    \section{Model}\label{model}
    \subsection{Non-Equilibrium Analysis of the Coevolutionary Model}\label{Non_Equilibrium}
    Here, we employ the non-equilibrium framework of the coevolutionary model including node--link--node triplets \cite{kargaran2021}. In this framework, the Hamiltonian of the system defines the interaction structure between nodes and links. Once the Hamiltonian is specified, the corresponding equations of motion can be derived from it, as in classical mechanics \cite{Goldstein}. In this formulation, the node variables $S_i(t)$ and the link variables $\sigma_{ij}(t)$ are treated as real-valued, time-dependent quantities. The dynamics are defined through effective interaction fields acting on nodes and links, leading directly to below differential equations:
    
    \begin{equation}\label{dynamic X}
        \begin{cases}
            dS_i/dt = \sum_j \sigma_{ij}\, S_j,\\
            d\sigma_{ij}/dt = S_i\, S_j.
        \end{cases}
    \end{equation}

    Here, the right-hand sides represent the interaction fields acting on nodes and links as determined by the Hamiltonian. The first equation shows that the evolution of a node is governed by the states of its neighbors weighted by the corresponding link variables, while the second equation shows that the evolution of a link is determined by the states of the nodes it connects.
    
    \subsection{Equilibrium Analysis of the Coevolutionary Model}\label{second_sub_model}  
    In coevolutionary models, the basic triad, two nodes and their connecting link, serves as the unit for defining the Hamiltonian\cite{kargaran2021}:  
    \begin{equation}\label{coevo_H}
    \mathcal{H}(G) = -\sum_{i<j} S_i \sigma_{ij} S_j .
    \end{equation}  
    where $G$ denotes a particular configuration of a network with $n$ nodes. In this formulation, the variables $S_i$ and $\sigma_{ij}$ are treated as discrete quantities representing the signs of the underlying dynamical variables in the long-time regime \cite{kulakowski,marvel2}. Configurations in which the system’s energy (defined as the value of the Hamiltonian) is minimized are referred to as balanced states, whereas all other configurations are imbalanced (Fig.~\ref{fig:main-fig1}). In the coevolutionary model, since both nodes and links can take values of either +1 or –1, two types of heaven states (global consensus) emerge. One corresponds to the situation in which nodes with positive opinions are in agreement: \tikzagreeone\ (type 1), and the other to the situation in which nodes with negative opinions are in agreement: \tikzagreetwo\ (type 2). The heaven state represents one of the system’s stable configurations, while another stable state is the bipolar configuration. The system becomes bipolar when one cluster adopts one type of heaven state and the other cluster adopts the other type, while the relations between the two clusters are characterized by disagreement.
            
    An analytical solution of this system has been obtained via the mean-field approach \cite{kargaran2021}. For node $i$, the Hamiltonian is:  
    \begin{equation*}\label{mean-field1}
    \mathcal{H}_i = - S_i \sum_{j \neq i} \sigma_{ij} S_j ,
    \end{equation*}  
    so that $\mathcal{H} = \mathcal{H}' + \mathcal{H}_i$, with $\mathcal{H}'$ independent of $S_i$.  The mean node value is \(p = \langle S_i \rangle = \sum_G S_i \mathcal{P}(G)\), with Boltzmann probability \(\mathcal{P}(G) = e^{-\beta \mathcal{H}(G)} / \mathcal{Z}\) and partition function \(\mathcal{Z} = \sum_G e^{-\beta \mathcal{H}(G)}\). Kargaran \etal \cite{kargaran2021} obtained the mean node value as  
    \begin{equation}\label{mean-link2-Xlayer}
        p = \tanh(\beta (n-1) q),
    \end{equation}  
    where $q$ is the node–link correlation. By decomposing the Hamiltonian into parts including node $i$, link $ij$, or both, one obtains:  
    \begin{equation}\label{mean-twostar-Xlayer}
        q = \tanh(\beta \tanh(\beta (n-1) q)).
    \end{equation}  
    where $n$ is the number of nodes in the network. In this equilibrium description, the system is characterized only by the signs of the variables, while their magnitudes are not explicitly considered.
            
    This admits one stable solution above the critical temperature ($T > T_c$), and two stable plus one unstable solution below it ($T < T_c$). As the social temperature decreases, the system experiences a second-order phase transition from disorder to order. Notably, the critical temperature of this transition is lower than that of the first-order phase transition observed in the thermal Heider balance model \cite{rabbani2019}.
    A comparison of these models reveals that in the Heider model, each link participates in more triangles, thereby amplifying the effect of sign changes. In contrast, while in the social Heider model the relation between two individuals depends on their mutual ties with a third party, in the coevolutionary model, it is determined directly by the opinions of the two individuals involved.\\

    \begin{figure}[t]
        \centering
        \includegraphics[width=1\linewidth]{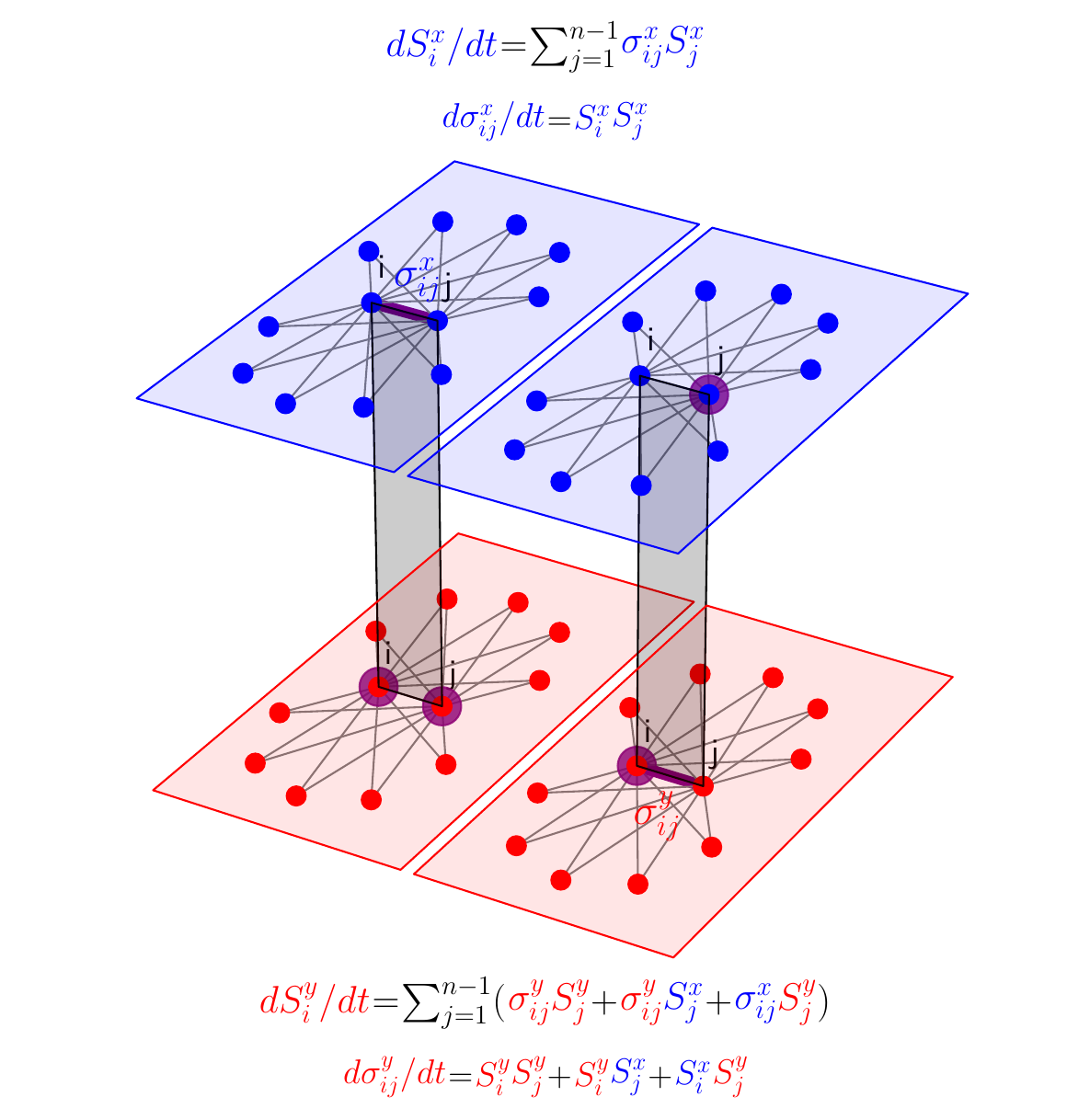}
        \caption{
        A schematic representation of the two-network system.  
        The independent network (blue) evolves autonomously, while the dependent network (red) evolves under its influence.  
        This one-way coupling highlights the asymmetry of the interaction.}
        \label{fig:main_fig2}
    \end{figure}
    
    \subsection{Two-Network Coevolutionary Model}\label{third_sub_model}
    We consider a system of two interacting networks with directed (asymmetric) coupling. The system consists of an independent network (denoted by $X$), which evolves autonomously, and a dependent (open) network (denoted by $Y$), whose dynamics are influenced by the state of the independent network. The two networks represent distinct populations of agents. Each network contains $n$ nodes and is taken to be fully connected, so that each node interacts with $n-1$ neighbors through node--link--node triplets.
    
    In the independent network ($X$), the dynamics are entirely governed by intra-network interactions. In this network, each link evolves based on the states of its two endpoints, while each node evolves under the aggregate interaction field generated by its connected neighbors. In contrast, the dependent network ($Y$) is influenced not only by its own internal structure, but also by the state of the independent network. The coupling is strictly one-way: the independent network evolves without feedback from the dependent network, while the dependent network explicitly contains interaction terms involving both its own variables and those of the independent network, as schematically shown in Fig.~\ref{fig:main_fig2}. For the independent network, the differential equations are those of Eq.~\eqref{dynamic X}, labeled by $x$. The differential equations are:
    
    \begin{widetext}
        \begin{equation}\label{dynamic Y}
            \text{dependent network} \,:\quad
            \left\{
            \begin{aligned}
            dS_{i}^{y}/dt &=
            \sum_{j=1}^{n-1} \left(
            \sigma_{ij}^{y} S_{j}^{y} +
            \sigma_{ij}^{y} S_{j}^{x} +
            \sigma_{ij}^{x}S_{j}^{y}  
            \right), \\
            d\sigma_{ij}^{y}/dt &= 
            S_{i}^{y} S_{j}^{y}
            +S_{i}^{y} S_{j}^{x}
            +S_{i}^{x} S_{j}^{y}.
            \end{aligned}
            \right.
        \end{equation}
    \end{widetext}
    This coupling ensures that the dependent network is partially driven by the independent network.
    
    To analyze the coupled dynamics, we introduce an auxiliary network $Z$. This transformation follows from combining the dynamical equations of the two networks, allowing the coupled system to be rewritten in a single set of variables. The network $Z$ is defined through $S_{i}^{z} = (S_{i}^{x} + S_{i}^{y})/2$ and $\sigma_{ij}^{z} = (\sigma_{ij}^{x} + \sigma_{ij}^{y})/2$. This transformation is exact and invertible. The auxiliary network $Z$ serves purely as a mathematical reparameterization of the coupled system, we have
    
    \begin{equation}\label{delta-E-z}
        \text{auxiliary network } Z:\quad
        \left\{
        \begin{aligned}
            dS_{i}^{z}/dt &= 2\sum_{j=1}^{n-1} \sigma_{ij}^{z} S_{j}^{z}, \\
            d\sigma_{ij}^{z}/dt &= 2S_{i}^{z} S_{j}^{z}.
        \end{aligned}
        \right.
    \end{equation}
    
    \section{Analysis}\label{analysis}
    \subsection{Analytical Framework Based on Mean-Field Theory}\label{first_sub_analysis}
    Our analysis now turns to the long-term behavior of the two-network coevolutionary model, where the system settles into its equilibrium states. We focus on the stationary configurations in which the fundamental quantities remain unchanged. To describe these equilibrium properties, we reformulate the dynamical equations in terms of the Hamiltonian. At the dynamical level, the node and link variables can be treated as continuous quantities evolving under the coevolutionary equations of motion. However, as shown in continuous formulations of structural balance~\cite{kulakowski,marvel2}, these systems exhibit a characteristic asymptotic behavior in which the magnitudes of the variables grow while their signs stabilize after a finite transient. In this regime, the system is effectively described by its sign structure, and the magnitudes become dynamically irrelevant. Accordingly, the variables $S_i$ and $\sigma_{ij}$ in the Hamiltonian formulation are defined as discrete quantities taking values in $\{\pm1\}$ for $X$ and $Y$ and $\{-1,0,1\}$ for $Z$, representing the stable signs of the underlying dynamical variables in the long-time limit.
    
    Using the mean-field approximation along with the Hamiltonians of the independent and auxiliary network, we can indirectly obtain the equilibrium solutions for the dependent network. Specifically, the Hamiltonians are given by:
    
    \begin{equation}\label{HamEqus}
    \left\{
    \begin{aligned}
    \mathcal{H}_x(G^{x}) &=-\sum_{i<j}S_{i}^{x}\,\sigma_{ij}^{x}\,S_{j}^{x},\\
    \mathcal{H}_z(G^{z}) &=-2\sum_{i<j}S_{i}^{z}\,\sigma_{ij}^{z}\,S_{j}^{z},
    \end{aligned}
    \right.
    \end{equation}
    where, to emphasize the distinction between the phase spaces of the networks, $G$ is labeled in the Hamiltonians. Here, $X$ denotes the independent network, while $Z$ is an auxiliary network introduced as an exact reparameterization of the coupled system.
        
    Inspired by the mean-field solution of the coevolutionary system (Section~\ref{second_sub_model}), we proceed to solve the two-network system. By applying the mean-field method to the independent and auxiliary networks, and using Eq.~\eqref{mean-link2-Xlayer} and~\eqref{mean-twostar-Xlayer}, we derive the corresponding equilibrium solutions for the dependent network we have
    \begin{equation}\label{definitions}
        \begin{aligned}
            p_y     &\equiv \langle S_i^y \rangle, \,\quad
            q_y     \equiv \langle S_i^y \sigma_{ij}^y \rangle, \\
            p_z     &\equiv \frac{p_x + p_y}{2} = \frac{ \langle S_i^x \rangle + \langle S_i^y \rangle }{2}, \\ 
            q_z     &\equiv \frac{q_x + q_y}{2} = \frac{ \langle S_i^x \sigma_{ij}^x \rangle + \langle S_i^y \sigma_{ij}^y \rangle }{2}, \\
            q_{xy}  &\equiv \frac{ \langle S_i^x \sigma_{ij}^y \rangle + \langle S_i^y \sigma_{ij}^x \rangle }{2}.
        \end{aligned}
    \end{equation}
    This analysis, as detailed in Appendix \ref{appendix:self-con-fun}, ultimately leads to the following self-consistent equations: one for the independent network and the other for the auxiliary network:
    
    \begin{equation}\label{selfcons 1}
    \left\{
    \begin{aligned}
    q_x &= f(q_x;\, n, T), \\
    q_z &= g(q_z;\, n, T).
    \end{aligned}
    \right.
    \end{equation}
    
    Next, supported by the equations obtained so far, we obtain the corresponding quantities for the dependent network:
    
    \begin{equation}
    p_y = 2p_z - p_x,
    \end{equation}
    
    \begin{equation}\label{q-y}
        \begin{aligned}
            q_{y} =
            \frac{
            4q_{z} - q_{x} - p_{x} q_{x} (8 p_{z}^3 - 12 p_{z}^2p_{x} + 6 p_{z}p_{x}^2 - p_{x}^3)
            }
            {
            1 + 2 p_{z} p_{x}^3 - p_{x}^4
            }.
        \end{aligned}
    \end{equation}
    The details of this part of our work can be found in Appendix \ref{appendix:q-follower-layer}. Moreover, based on the calculations presented in Appendix \ref{appendix:E-follower-layer}, the energy of the dependent network can be expressed as follows:
    
    \begin{equation}
    \begin{aligned}
    E_{y} &= 
    (8E_{z} - E_{x}) + p_{x} (p_{x}^3 p_{y} q_{y} + p_{y}^3 p_{x} q_{x} + q_{y}) \\
    &\qquad\qquad\qquad\qquad+ p_{y} (p_{x}^3 p_{y} q_{y} + p_{y}^3 p_{x} q_{x} + q_{x}).
    \end{aligned}
    \end{equation}

    \begin{figure}[t]
        \centering
        \includegraphics[width=\linewidth]{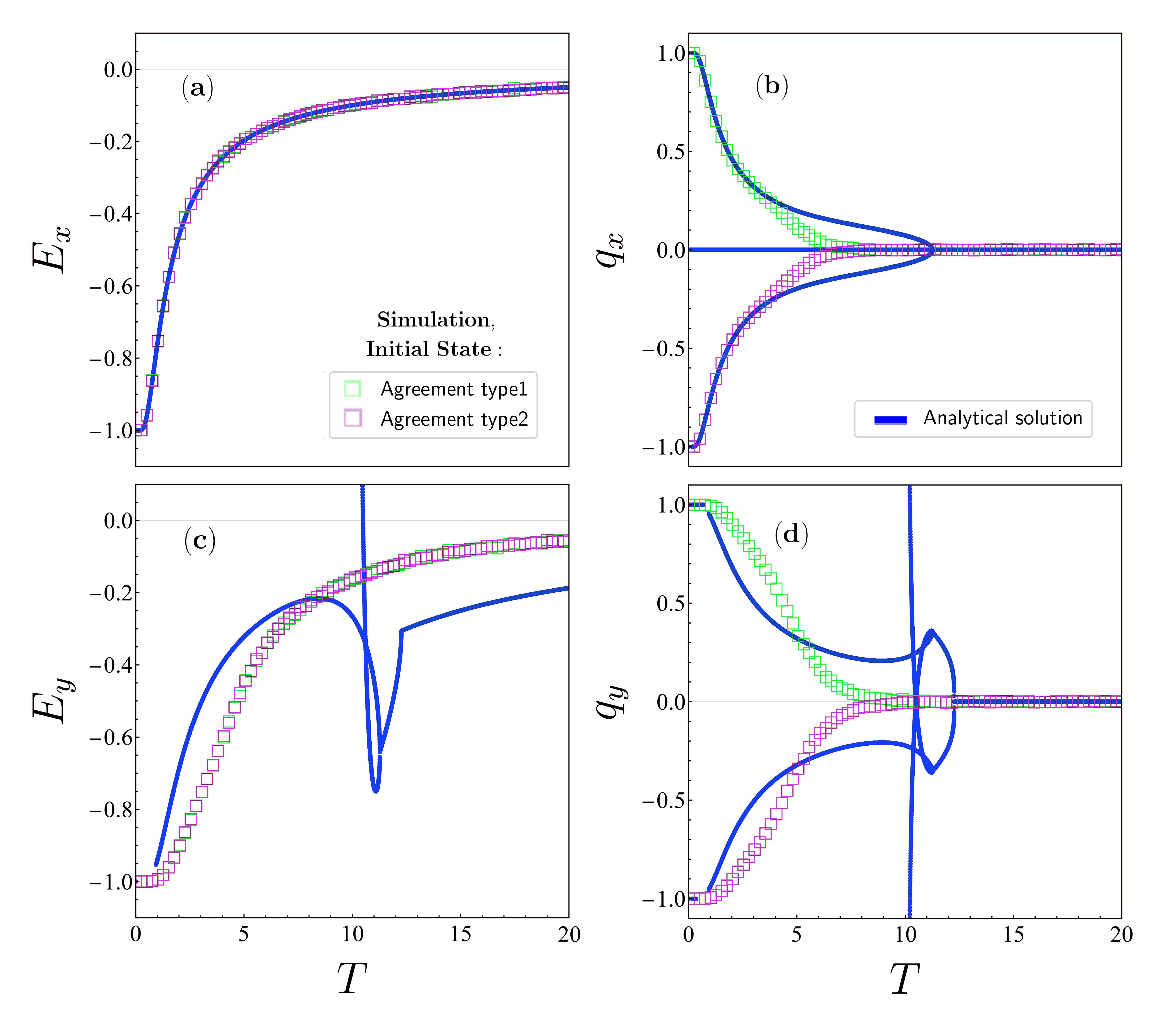}
        \caption{Comparison between theoretical predictions and Monte Carlo simulations for the temperature dependence of energy and order parameter in both networks. Panels~(a) and (b) show the energy and order parameter of the independent network (X), respectively, while panels~(c) and (d) display the same quantities for the dependent network (Y). Solid lines represent the analytical results obtained from self-consistent equations, and symbols correspond to simulation data. Green and purple squares denote agreement type~1:~\tikzagreeone\ and agreement type~2:~\tikzagreetwo, respectively. The independent network exhibits a lower critical temperature than the dependent network. The system contains $n=128$ nodes.}
        \label{fig:main_fig3}
    \end{figure}
    
    \begin{figure}[t]
        \centering
        \includegraphics[width=1\linewidth]{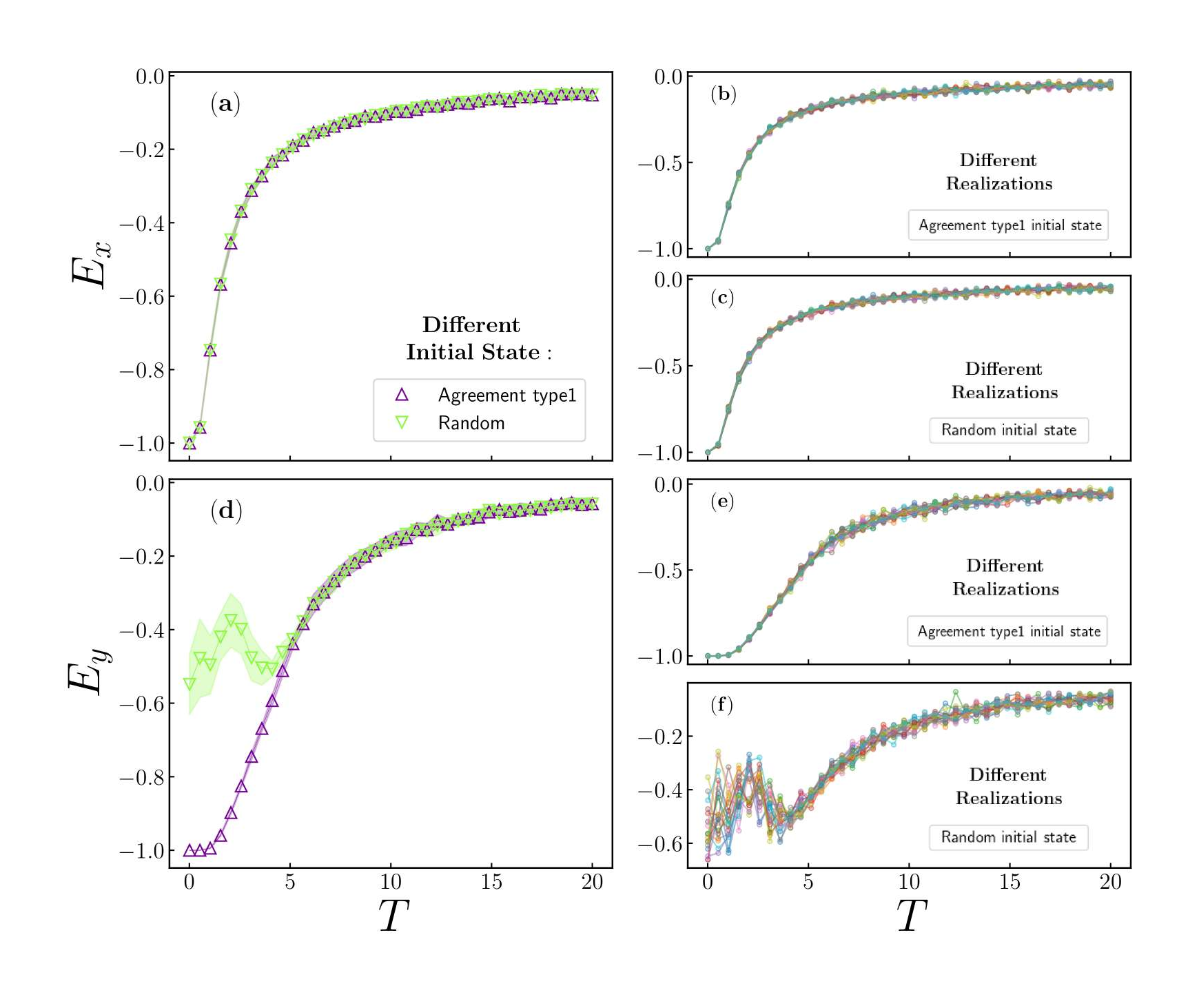}
        \caption{The results of the simulations for the independent and dependent networks. Independent network: panel~(a) shows the energy for all positive and random initial states (magenta/green), with panels~(b)–(c) depicting representative realizations. Dependent network: panel~(d) shows the energy for all positive and random initial states, with panels~(e)–(f) depicting representative realizations. Solid lines denote ensemble averages and shaded area indicate one standard deviation. Results are based on 20 realizations with $n=128$ nodes.}
        \label{fig:main_fig4}
    \end{figure}
    \subsection{Simulations}\label{second_sub_analysis}
    To validate the analytical results, we performed Monte Carlo simulations using the Metropolis algorithm, which samples configurations according to the Boltzmann distribution. We directly simulate the coupled two-network system consisting of an independent network ($X$) and a dependent network ($Y$). At each step, a single degree of freedom—either a node or a link—is selected randomly from the combined set. A node is chosen with probability $p = \nicefrac{n}{n + n(n-1)/2}$ and a link with probability $1-p$, ensuring uniform sampling over all variables, consistent with standard coevolutionary models~\cite{kargaran2021}. The independent network ($X$) evolves autonomously, with energy changes determined solely by its internal triplet interactions. In contrast, the dependent network ($Y$) evolves under asymmetric coupling: its energy change includes contributions from both internal triplets within $Y$ and mixed triplets involving variables from $X$ and $Y$, as defined by Eqs.~\eqref{dynamic X} and \eqref{dynamic Y}. As a result, updates in $Y$ must simultaneously satisfy balance constraints within its own structure and those imposed by the $X$ network. The simulations therefore implement the original $(X,Y)$ system directly; the auxiliary network $Z$ is used only in the analytical treatment and is not simulated.

    Figure~\ref{fig:main_fig3} compares the Monte Carlo simulations with the mean-field predictions for the temperature dependence of the energy and order parameter in both networks. Panels~(a) and~(b) show the energy and order parameter of the independent network~($X$), respectively, while panels~(c) and~(d) display the corresponding quantities for the dependent network~($Y$). The simulations are initialized in homogeneous agreement-type states, where all agents experience approximately the same effective local field. Under these conditions, the assumptions of the mean-field approximation remain valid, leading to good agreement between the analytical self-consistent solutions (solid lines) and the simulation results (symbols) for both networks. The results also show that the independent network undergoes the transition at a lower critical temperature than the dependent network. In contrast, random initial conditions produce spatially heterogeneous bipolar configurations with nonuniform local fields, violating the homogeneity assumption of the mean-field framework; consequently, reliable mean-field predictions are not expected in that regime.
    
\begin{figure}[t]
        \centering
        \includegraphics[width=\linewidth]{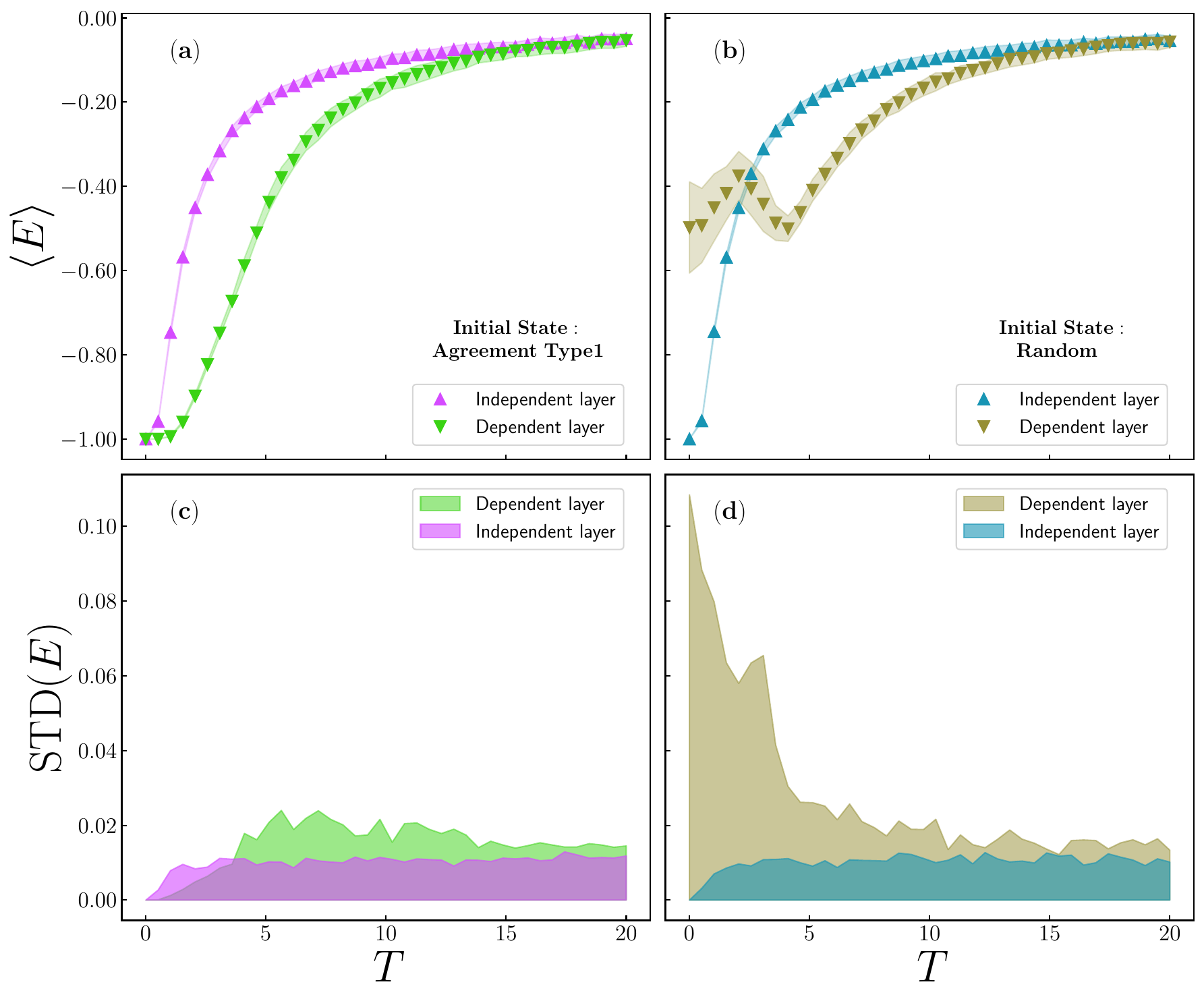}
        \caption{Panel~(a) Mean energy $\langle E \rangle$ of the independent and dependent networks for an all-positive initial state. Panel~(b) same for random initial conditions. Panels~(c) and (d) show energy fluctuations. The system contains $n=128$ nodes.}
        \label{fig:main_fig5}
    \end{figure}

    The behavior under different initial conditions is summarized in Figs.~\ref{fig:main_fig4} and \ref{fig:main_fig5}. In Fig.~\ref{fig:main_fig4}, panels (a)–(c) show the temperature dependence of the mean energy for the independent network ($X$), while panels (d)–(f) show the corresponding results for the dependent network ($Y$). In Fig.~\ref{fig:main_fig5}, the upper panels display the mean energy and the lower panels show the standard deviation, allowing a direct comparison between the two networks. For agreement-type initial conditions [Fig.~\ref{fig:main_fig4}(a,b) for $X$ and Fig.~\ref{fig:main_fig4}(d,e) for $Y$], both networks evolve toward the same ordered state: the energy approaches the fully balanced value as $T \to 0$ and fluctuations remain small, consistent with mean-field predictions. In contrast, for random initial conditions [Fig.~\ref{fig:main_fig4}(c) for $X$ and Fig.~\ref{fig:main_fig4}(f) for $Y$], the behavior changes qualitatively. As seen in Fig.~\ref{fig:main_fig4}(c), the independent network ($X$) still relaxes toward a balanced (typically bipolar) state with small fluctuations. However, Fig.~\ref{fig:main_fig4}(f) shows that the dependent network ($Y$) does not reach the fully balanced state and exhibits persistent variability even at low temperatures. This occurs because $Y$ is subject to two competing influences: its own internal balance dynamics and the external structure imposed by $X$, which generally favor different configurations. As a result, $Y$ remains effectively frustrated. This frustration is further confirmed in Fig.~\ref{fig:main_fig5} (lower panels), where the standard deviation of the energy in $Y$ is significantly larger than in $X$, indicating strong realization-to-realization variability and the presence of multiple competing attractors. This behavior is absent in the independent network and is a direct consequence of the asymmetric coupling.
    
    \section{CONCLUSIONS} \label{sec:conclusion}
    We consider a system of two interacting networks with directed (asymmetric) coupling, consisting of an independent network and a dependent (open) network whose dynamics are influenced by the former. Our study of coevolutionary balance in open networks leads to three principal conclusions:
    \begin{itemize}
        \item Persistence of Imbalance states: Unlike closed systems, open networks do not necessarily reach complete structural balance; imbalanced states can survive metastable even below the critical temperature.
        \item Ensemble Dependence and Uncertainty: The final state of the system is not unique but depends on the ensemble, indicating a fundamental uncertainty in predicting balance outcomes.
        \item Enhanced Disorder Tolerance: Open systems are more robust to thermal noise, as evidenced by a higher critical temperature compared to their closed counterparts.
    \end{itemize}
    
    These results, validated by both mean-field theory and simulation, highlight the necessity of incorporating external influence, here represented by the directed coupling from the independent network, which acts as an effective environment for the dependent network, for a realistic model of balance in real-world networks.
    
    \appendix
    \begin{appendices}
    \section{Analytical Derivation of the Equations for the Auxiliary Network $Z$}
    
    The network $Z$ is an auxiliary network introduced through the transformation between the independent network $X$ and the dependent network $Y$. It is a reparameterization introduced to enable analytical treatment.
    
    \subsection{Self-Consistent Equations}\label{appendix:self-con-fun}
    
    We can decompose the Hamiltonian of the auxiliary network $Z$ into two parts: one that includes the element $i$ ($\mathcal{H}_{i}^z$) and another part that contains all the remaining terms ($\mathcal{H}'^z$).
    \begin{align*}
        \mathcal{H}_z(G^{z}) &=-2\sum_{i<j}S_{i}^{z}\,\sigma_{ij}^{z}\,S_{j}^{z},  
    \end{align*}
    
    \begin{align}\label{H'}
        \mathcal{H}^z &= \mathcal{H}_i^z + \mathcal{H}'^z.
    \end{align}
    
    Similarly, the Hamiltonian can also be decomposed into two parts in another way:
    \begin{align}\label{H''}
        \mathcal{H}^z &= \mathcal{H}_{ij}^z + \mathcal{H}''^z,
    \end{align}
    where \(\mathcal{H}_{ij}^z\) for the auxiliary network $Z$ includes terms involving the elements \(i\), \(j\), or both, and \(\mathcal{H}''^z\) contains all other remaining terms. To further clarify the structure of each term in the Hamiltonian decomposition, we explicitly present the expressions for both $\mathcal{H}_i^z$ and $\mathcal{H}_{ij}^z$ as follows:
    
    \begin{equation}
    \left\{
    \begin{aligned}
        \mathcal{H}^z_i   &= -2 S_i^z \sum_{j \neq i} \sigma_{ij}^z S_j^z, \\
        \mathcal{H}^z_{ij} &= -2 S_i^z \sigma_{ij}^z S_j^z -2 S_j^z \sum_{k \neq i,j} \sigma_{jk}^z S_k^z.
    \end{aligned}
    \right.
    \end{equation}
    
    Eq.~\eqref{coevo_H}--\eqref{mean-twostar-Xlayer}, as introduced in Ref.~\cite{kargaran2021}, are valid for the independent network, so we labeled it with $x$. Now, to calculate the equations of the auxiliary network $Z$, we follow the same method. However, we must keep in mind that, unlike the independent network, the variables of $Z$ may also be zero because they are defined as averages of the corresponding variables in $X$ and $Y$. The mean value of the nodes is written as follows:
    
    \begin{align}
        &p_{z} = \langle S_{i}^z \rangle = \sum_{G^z} S_i^z \, \mathcal{P}(G^z) \label{eq:mean_spin} \\[6pt]
        &\mathcal{P}(G^z) = \frac{e^{-\beta \mathcal{H}^{z}(G^z)}}{\mathcal{Z}^z} \notag \\[6pt]
        &\mathcal{Z}^z = \sum_{G^z} e^{-\beta \mathcal{H}^{z}(G^z)} \notag \\[6pt]
    \end{align}   
    where \(\mathcal{P}(G^z)\) is the Boltzmann weight and \(\mathcal{Z}^z\) is the partition function, both defined in the phase space of the auxiliary network $Z$. Next, by integrating over the phase space and substituting Eq.~\eqref{H'} into Eq.~\eqref{eq:mean_spin}, we obtain:
    
    \begin{align}
        p_{z} &= \frac{1}{\mathcal{Z}^z} \sum_{S^z \neq S_{i}^z} e^{-\beta \mathcal{H}'^{z}} 
        \sum_{S_{i}^z \in \{ 0, \pm 1 \}}
      S_{i}^z e^{-\beta \mathcal{H}_{i}^{z}} \notag \\[6pt]
         &= \frac{e^{2\beta (n-1) q_z} - e^{-2\beta (n-1) q_z}}{e^{2\beta (n-1) q_z} +
        e^{-2\beta (n-1) q_z} + 1} \\[10pt]
         &=  \frac{2 \sinh\left(2 (n - 1) q_z \beta \right)}{1 + 2 \cosh\left(2 (n - 1) q_z \beta \right)}
         .\label{eq:analytic_solution}
    \end{align}
    
    Similarly, by defining the node-link correlation for the auxiliary network $Z$ and performing the integration over the corresponding phase space along with Eq.~\eqref{H''}, we obtain:
    \begin{center}
    \begin{minipage}{0.48\textwidth}
    \begin{equation}\label{mean-field3}
    q_{z} = \langle{\sigma_{ij}^z S_{j}^z}\rangle = \sum_{G^z} \sigma_{ij}^z S_{j}^z \mathcal{P}(G^z)
    \end{equation}
    \end{minipage}
    
    \hfill
    \begin{minipage}{0.48\textwidth}
    \begin{equation}\label{kula-dynamic}
    q_z =
    \frac{
        \displaystyle
        \sum_{S^z,\, \sigma^z \neq S_{j}^z ,\, \sigma_{ij}^z} e^{-\beta \mathcal{H}''^z}
        \sum_{\substack{
            S_j^z,\, \sigma_{ij}^z \in \{ 0, \pm 1 \}
        }}
        \sigma_{ij}^z S_j^z\, e^{-\beta \mathcal{H}_{ij}^z}
    }{
        \displaystyle
        \sum_{S^z,\, \sigma^z \neq S_{j}^z ,\, \sigma_{ij}^z} e^{-\beta \mathcal{H}''^z}
        \sum_{\substack{
            S_j^z,\, \sigma_{ij}^z \in \{ 0, \pm 1 \}
        }}
        e^{-\beta \mathcal{H}_{ij}^z}
    }
    \end{equation}
    \end{minipage}
    \end{center}
    
    \begin{widetext}
        \begin{equation}
        q_z =
        \frac{
            2 \Bigg(
                -\sinh\left(
                    2 (n - 2) q_z \beta - \frac{4 \beta \sinh\left(2 (n - 1) q_z \beta\right)}{1 + 2 \cosh\left(2 (n - 1) q_z \beta\right)}
                \right)
                +
                \sinh\left(
                    2 (n - 2) q_z \beta + \frac{4 \beta \sinh\left(2 (n - 1) q_z \beta\right)}{1 + 2 \cosh\left(2 (n - 1) q_z \beta\right)}
                \right)
            \Bigg)
        }{
            3
            + 2 \cosh\left(2 (n - 2) q_z \beta\right)
            + 2 \cosh\left(
                2 (n - 2) q_z \beta - \frac{4 \beta \sinh\left(2 (n - 1) q_z \beta\right)}{1 + 2 \cosh\left(2 (n - 1) q_z \beta\right)}
            \right)
            + 2 \cosh\left(
                2 (n - 2) q_z \beta + \frac{4 \beta \sinh\left(2 (n - 1) q_z \beta\right)}{1 + 2 \cosh\left(2 (n - 1) q_z \beta\right)}
            \right)
        }
        \end{equation}
    \vspace{1em}    
    \end{widetext}
    
    Therefore, we have obtained two self-consistent equations: one for the independent network $X$ and one for the auxiliary network $Z$:   
    \begin{equation}\label{selfcons}
        \left\{
        \begin{aligned}
            q_x &= f(q_x;\, n, T), \\
            q_z &= g(q_z;\, n, T).
        \end{aligned}
        \right.
    \end{equation}
    
    \subsection{Derivation of the Energy Equation}
    With the background of the equations $\displaystyle \mathcal{H}_z(G^{z}) =-2\sum_{i<j}S_i^z\, \sigma_{ij}^z \, S_j^z$ \eqref{HamEqus} and $\displaystyle \mathcal{H}^z= \mathcal{H}_{ij}^z + \mathcal{H}''^z$ \eqref{H''}, we can write:
    \begin{equation}\label{mean-field1}
        \mathcal{H}_{i,j}^z= - S_{i}^z{\sum_{l\neq{i,j}}\sigma_{il}^z \, S_{l}^z}- S_{j}^z{\sum_{l\neq{i,j}}\sigma_{jl}^z \, S_{l}^z} - S_{i}^z \, \sigma_{ij}^z \, S_{j}^z.
    \end{equation}
    The expected value of $\langle S_i^z \sigma_{ij}^z S_j^z \rangle$ is given by the sum over all possible configurations, where each term is the product of its corresponding value and its Boltzmann weight:
    
    \begin{equation}
        \langle S_{i}^z \sigma_{ij}^z S_{j}^z \rangle = 
        \frac{1}{Z^z} \sum_{S_{i}^z, \sigma_{ij}^z, S_{j}^z} 
        (S_{i}^z \sigma_{ij}^z S_{j}^z) \, e^{-\beta H^z\,(S_{i}^z,\, \sigma_{ij}^z,\, S_{j}^z)}.
    \end{equation}
    By definition, the partition function $Z^z$ is the sum over all possible configurations of $S_i^z$, $\sigma_{ij}^z$, and $S_j^z$. After simplifications, the energy of the auxiliary network $Z$ is obtained as follows:
    
    \vspace{1em}
    \begin{widetext}        
    \begin{equation}
    E_z = -\frac{8 \cosh^{2}\left(\lambda \beta\right) \sinh\left(2 \beta\right)}{
        5 + 4 \cosh(2 \beta)
        + 12 \cosh\left(\lambda \beta\right)
        + 2 \cosh\left(2\lambda \beta\right)
        + 2 \cosh\left(2 \left(1 + \lambda \right) \beta \right)
        + 2 \cosh\left(2 \left(1 + 4 q_z - 2 N q_z \right) \beta \right)
    } ,
    \end{equation}
    \vspace{1em}
    where \( \lambda = 2(N - 2)q_{z} \).
    \end{widetext}
        
    \section{Analytical Derivation of the Equations for the Dependent Network}
    
    The dependent network $Y$ is not solved independently. Instead, its equilibrium quantities are reconstructed from the independent network $X$ and the auxiliary network $Z$, using the exact transformation between $(X,Y)$ and $(X,Z)$.
    
    \subsection{Self-Consistent Equation}\label{appendix:q-follower-layer}
    Using the equations of the independent network $X$ and the auxiliary network $Z$, we derive the equations for the dependent network $Y$. First, the mean values of nodes in the networks are related as follows:
    \begin{align}
        p_z &= \frac{p_{x} + p_{y}}{2}, \nonumber \\
        p_y &= 2p_z - p_x. \label{py_definition}
    \end{align}
    
    Next, we turn our attention to the node-link correlations:
    \begin{align}
        q_{z}   &= \langle \sigma^z_{ij} S_{j}^z\rangle \nonumber \\[6pt]
                &=  \left( \langle \sigma_{ij}^x S_{j}^x\rangle  
                        + \langle \sigma_{ij}^y S_{j}^y\rangle  
                        + \langle \sigma^y_{ij} S_{j}^x\rangle
                        + \langle \sigma_{ij}^x S_{j}^y\rangle \right)/4 \nonumber \\[6pt]
                &= \left( q_{x} + 2 q_{xy} + q_{y}\right)/4 \label{eq:qz} .\\[6pt]
    \end{align}
    
    To evaluate $q_{xy}$, we explicitly expand each mixed term in terms of the mean value of the node and the correlation between the node and the link. Here we use the mean-field factorization, so higher-order mixed averages are approximated by products of lower-order averages and correlations.
    \begin{equation}
        \left\{
        \begin{aligned}
            \langle S_{i}^x\,\sigma_{ij}^y\rangle 
            &=\langle S_{i}^x\,S_{i}^x\,S_{i}^x\,S_{i}^y\,S_{i}^y\,\sigma_{ij}^y\rangle
            = p_{x}^2\, p_{x}\,p_{y}\, q_{y} ,\\[6pt]
            \langle S_{i}^y\,\sigma_{ij}^x\rangle &=\langle S_{i}^y\,S_{i}^y\,S_{i}^y\,S_{i}^x\,S_{i}^x\,\sigma_{ij}^x\,\rangle
            = p_{y}^2\, p_{y}\,p_{x}\, q_{x} .\\                       
        \end{aligned}
        \right.
    \end{equation}            
    
    Substituting these expressions into Eq.~\eqref{definitions}, we obtain the following compact form for $q_{xy}$:
    
    \begin{equation}
        \begin{aligned}
            q_{xy} &= (p^2_{x}\, p_{x}\, p_{y}\, q_{y} + p^2_{y}\, p_{y}\, p_{x}\, q_{x})/2 \\[6pt]
                   &= (p_{x}^3\, p_{y}\, q_{y} + p_{y}^3\, p_{x}\, q_{x})/2 .
        \end{aligned}
    \end{equation}  
    
    By substituting $q_{xy}$ into Eq.~\eqref{eq:qz}, the correlation $q_z$ in the auxiliary network $Z$ can be expressed as:
    
    \begin{equation}
        \begin{aligned}
            &q_{z} =(q_{x} + p_{x}^3 \,p_{y} q_{y} + p_{y}^3\, p_{x} q_{x} + q_{y})/4.
        \end{aligned}
    \end{equation} 
    
    Rearranging this equation allows us to solve for $q_y$ in terms of $q_x$, $q_z$, and $p_x, p_y$:
    
    \begin{equation}
        \begin{aligned}
            q_{y} =
                \frac{
                    4q_{z} - q_{x} - p_{y}^3 \,p_{x} q_{x}
                    }
                    {
                    1 + p_{x}^3 p_{y}
                    }.
        \end{aligned}   
    \end{equation}
    Finally, substituting the relation $p_y = 2p_z - p_x$ from Eq.~\eqref{py_definition}, we obtain:
    
    \begin{equation}
        \begin{aligned}
            q_{y} =
                \frac{
                    4q_{z} - q_{x} - (2 p_{z} - p_{x})^3 \,p_{x} q_{x}
                    }
                    {
                    1 + p_{x}^3 \,(2 p_{z} - p_{x})
                    },
        \end{aligned}   
    \end{equation}     
    which, after expansion and simplification, leads to the final compact form:
    \begin{equation}\label{q-y}
        \begin{aligned}
            q_{y} =
                \frac{
                    4q_{z} - q_{x} - p_{x} q_{x} (8 p_{z}^3 - 12 p_{z}^2p_{x} + 6 p_{z}p_{x}^2 - p_{x}^3) 
                    }
                    {
                    1 + 2 p_{z} p_{x}^3 - p_{x}^4
                    }.
        \end{aligned}   
    \end{equation}  
           
    \subsection{Energy of the Dependent Network}\label{appendix:E-follower-layer}
    We start by considering the Hamiltonian of the auxiliary network $Z$, where the contributions of individual node-link-node triplets can be expressed in terms of the corresponding elements of the independent and dependent networks:
    
    \begin{align*}
        \mathcal{H}_z(G^{z}) &=-2\sum_{i<j}S_{i}^{z}\,\sigma_{ij}^{z}\,S_{j}^{z},    
    \end{align*}
    as we know, the auxiliary network $Z$ is defined from the independent and dependent networks as:
    \begin{equation}            
    \begin{aligned}
        S_i^z      = \frac{  S_i^x  + S_i^y  }{2}, \, \quad
        \sigma_{ij}^z   = \frac{ \sigma_{ij}^x  +  \sigma_{ij}^y   }{2}.
    \end{aligned}
    \end{equation}
    
    So we can write:
    \begin{equation}\label{HamEqus2}
    \begin{aligned}
        S_i^z \sigma_{ij}^z S_j^z 
        &= \frac{(S_i^x + S_i^y)(\sigma_{ij}^x + \sigma_{ij}^y)(S_j^x + S_j^y)}{8} \\
        &= \frac{1}{8} \big(
            S_i^x \sigma_{ij}^x S_j^x 
          + S_i^x \sigma_{ij}^x S_j^y 
          + S_i^x \sigma_{ij}^y S_j^x \\
        &\quad + S_i^x \sigma_{ij}^y S_j^y 
          + S_i^y \sigma_{ij}^x S_j^x 
          + S_i^y \sigma_{ij}^x S_j^y \\
        &\quad + S_i^y \sigma_{ij}^y S_j^x 
          + S_i^y \sigma_{ij}^y S_j^y
        \big)
        ,
    \end{aligned}
    \end{equation}
    which allows us to separately evaluate the contributions from mixed terms.  
    Using the mean-field approximation, we can compute the averages of these terms as follows:
    
    \begin{equation}\label{rateEqZ}
    \left\{
    \begin{aligned}
        &\langle S_i^x \sigma_{ij}^y S_j^y \rangle = \langle S_i^x \rangle \langle \sigma_{ij}^y S_j^y \rangle = p_x q_y,\\
        &\langle S_i^y \sigma_{ij}^x S_j^x \rangle = \langle S_i^y \rangle \langle \sigma_{ij}^x S_j^x \rangle = p_y q_x,\\
        &\langle S_i^x \sigma_{ij}^x S_j^y \rangle + \langle S_i^x \sigma_{ij}^y S_j^x \rangle = 2 p_x q_{xy},\\
        &\langle S_i^y \sigma_{ij}^x S_j^y \rangle + \langle S_i^y \sigma_{ij}^y S_j^x \rangle = 2 p_y q_{xy}.
    \end{aligned}
    \right.
    \end{equation}
    
    Finally, the following expression completes the derivation of the dependent network's energy based on the contributions from both independent and auxiliary networks and their correlations:
    
    \begin{align}
    E_y &= (8 E_z - E_x) \nonumber\\
        &\quad + p_x (p_x^3 p_y q_y + p_y^3 p_x q_x + q_y) \nonumber\\
        &\quad + p_y (p_x^3 p_y q_y + p_y^3 p_x q_x + q_x).
    \end{align}
    \end{appendices}
    
        \bibliographystyle{apsrev4-2}
        \bibliography{reference_CBTN}       

\end{document}